\documentclass[10pt]{iopart}
\usepackage[dvips]{graphics}
\usepackage{epsfig}
\def\prep{{\it Preprint}}
\def\thp	{$\Theta^+$}
\def\ximm	{$\Xi^{--}_{3/2}$}
\def\xipp	{$\bar{\Xi}^{++}_{3/2}$}
\def\xim	{$\Xi^{-}$}
\def\xip	{$\bar{\Xi}^{+}$}

\def\xis	{$\Xi^{*}$}

\def\xisz	{$\Xi(1530)^0$}
\def\axisz	{$\bar{\Xi}(1530)^0$}

\def\pip	{$\pi^+$}
\def\pim	{$\pi^-$}
\def\ks		{K$^0_S$}
\def\kp         {K$^+$}
\def\km         {K$^-$}

\def\lam	{$\Lambda$}
\def\alam	{$\bar{\Lambda}$}
\def\sol        {{\it c}}
\def\solq       {{\it c$^2$}}

\begin{document}
\title[Search for pentaquarks in HERA-B]
{Search for \thp\ and \ximm\ pentaquarks in HERA-B}

\author{K T Kn\"opfle\dag, M Zavertyaev\dag\ and T \v Zivko\ddag \\  
                for the HERA-B collaboration}
\address{\dag\ Max-Planck-Institut f\"ur Kernphysik,
          D-69117 Heidelberg, Germany}
\address{\ddag\ Institut Jo\v zef Stefan, 1001 Ljubljana, Slovenia}

\ead{Karl-Tasso.Knoepfle@mpi-hd.mpg.de} 

\begin{abstract}
A search for \thp(1540) and \ximm(1860) pentaquark candidates has been
performed in proton-induced reactions on  C, Ti and W targets at 
$\sqrt{s} = 41.6$~GeV studying the p\ks\ resp. \xim\pim\
and \xim\pip\  (and charge conjugated (c.c.)) decay channels at mid-rapidity.
With sensitivities of Br$\times$d$\sigma$/dx$_F~<~5$ to 25~$\mu$b/nucleon, 
we find no evidence for narrow pentaquark peaks in any of the studied final 
states.
Preliminary values for the upper limit of relative yield ratios at mid-rapidity
are \thp(1540)~/~\lam(1520)~$<$~0.02,
Br$\times$\ximm(1862)~/~\xisz~$<$~0.077, and  
Br$\times$\xipp(1862)/\axisz~$<$~0.058
at 95\% CL. 
\end{abstract}


\section{Introduction}
\label{sec:intro}

Recently, evidence for an exotic S~=~+1 baryon has been reported by several experiments
using incident beams of real and quasi-real photons, kaons, and (anti-)neutrinos \cite{leps}. 
With a significance
of 4.4 to 7$\sigma$, they find in the p\ks\ or n\kp\ channels a resonance with a mass between 1526 
and 1542~MeV/\solq\ and with a width of typically $<~20$~MeV dominated by the experimental resolutions. 
Best constraints on width come  from a re-analysis of kaon scattering and exclude a 
value larger than 5~MeV/\solq\ \cite{cah03}. 
This new state is tentatively  identified with the \thp\ pentaquark ($uudd\bar{s}$) baryon,  
which has been predicted \cite{dia97} within a chiral soliton model to reside at 1530 MeV/\solq , 
to have a width of less than 15~MeV/\solq , and to decay into the KN channels. Other model 
interpretations including correlated quark pairs are discussed in \cite{jaf03}.
The \thp\ is member of an antidecuplet exhibiting two further exotic states, the \ximm\  
($ddss\bar{u}$) and the $\Xi^+_{3/2}$ ($uuss\bar{d}$).
A candidate for the \ximm\ 
has been found by the NA49 collaboration in the \xim\pim 
channel at 1862 MeV/\solq\ with a width of less than 18 MeV/\solq\ \cite{na49}. 
At the same mass, evidence for a new resonance is also seen in the \xim\pip +\,c.c. channels. 
For establishing the existence and character of the new resonances beyond any doubt, high statistics 
mass spectra and the measurements of spin, parity, width and cross sections are needed. 
Exploring the production mechanism of pentaquark states can  lead to a better understanding
of non-perturbative QCD, and relative particle yields like \thp / \lam\ or \thp /\lam(1520) can provide
new insight in the dynamics of production mechanisms. Statistical hadronization models
predict these ratios to be $>$0.02 resp. $>$0.4 and energy-dependent \cite{ran03,let03,bec03}. 
Taking advantage of a huge sample of minimum bias data from proton-nucleus collisions, HERA-B can
considerably contribute to all these topics. 

\section{Experiment and data sample}
\label{sec:exp}

\begin{figure}[h]
\begin{center}
\includegraphics[width=11cm]{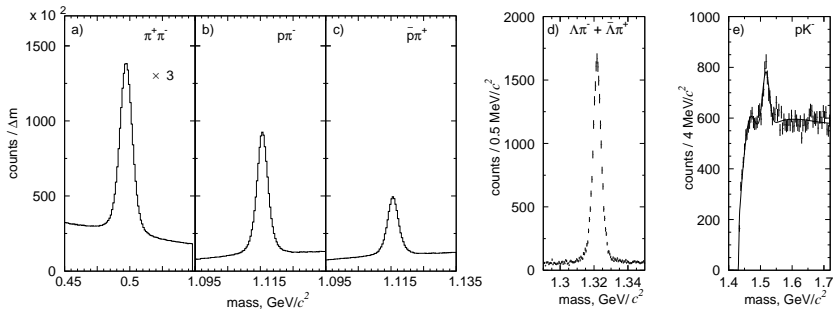}
\end{center}
\caption{\label{d1} Signals from decays of 
                 a) K$^0_S\rightarrow \pi^+\pi^-$,
		 b) $\Lambda\rightarrow$p$\pi^-$, 
		 c) $\bar{\Lambda}\rightarrow\bar{p}\pi^+$,
		 d) \xim$\rightarrow$\lam\pim\,and \xip$\rightarrow$\alam\pip , and 
		 e) $\Lambda(1520)\rightarrow$pK$^-$. $\Delta$m=1.0\,/\,0.4\,MeV/\solq\ in a)\,/\,b),\,c).
}		  
\end{figure}

HERA-B is a fixed target experiment at the 920 GeV proton storage ring of DESY. 
It is a forward magnetic spectrometer with a large acceptance, a high-resolution vertexing and 
tracking system and good particle identification \cite{abt03}.
The present study uses a sample of more than 200 million 
minimum bias events which was taken at mid-rapidity ($x_F \sim 0$) on carbon, titanium and tungsten targets 
during the 2002/03 run period.
For this analysis the information from the silicon vertex detector, the main tracking system 
and the ring-imaging Cherenkov counter (RICH) was used.
With standard techniques described in \cite{abt03}, signals from \ks~$\rightarrow$~\pip\pim ,
\lam~$\rightarrow$~p\pim\ and \alam~$\rightarrow$~$\bar{\rm p}$\pip\ decays can be identified above
small backgrounds (Figure \ref{d1}a,b,c).
Similar clean signals from \xim~$\rightarrow$~\lam\pim\ and c.c. decays (Figure \ref{d1}d) are obtained
by requesting that none of the decay products but only the \xim\ itself must point to the 
primary vertex. The measured mass resolutions  for \ks , \lam\ and \xim (\xip) signals of 4.9~MeV,
1.6~MeV, and 2.6~MeV are about 20\% larger than the values from Monte Carlo simulations.   
The number of observed \ks , \lam +\alam\ and \xim +\xip\  events totals about $3.4\times 10^6$, 
$1.4\times 10^5$, and $1.9\times 10^4$, respectively.   

\section{Search for \thp~$\rightarrow$~p~+~\ks\ decays}
\label{sec:searchpk}

\begin{figure}[h]
\centering
\includegraphics[width=14cm]{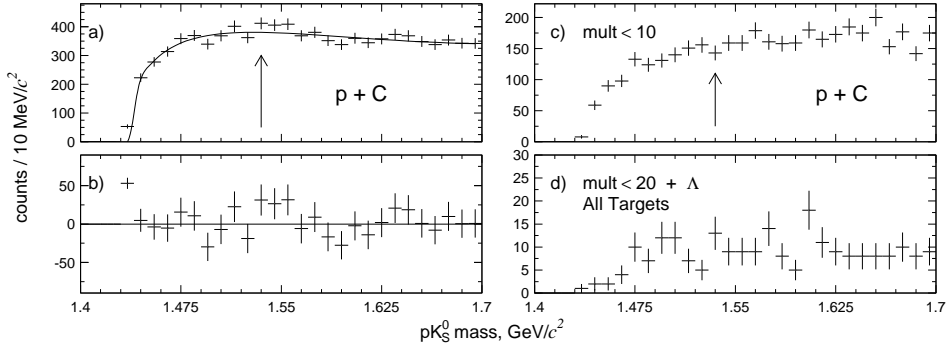}
\caption{\label{d2} 
The pK$^0_S$ invariant mass distributions: a) data from the p+C collisions with 
background (continuous line) determined from event mixing; b) as a) but with the background
subtracted; c) as a) but requiring a track multiplicity of $<10$. d) data from all targets
C, Ti, W requiring a track multiplicity of $<20$ and a $\Lambda$ particle in the event.
Arrows mark the mass of 1540~MeV/\solq .}
\end{figure}

The large statistics of \ks\ events allows to apply very strict criteria for proton identification 
by the RICH. 
The cut in the proton likelihood of $>0.95$ implies that the probability to identify other particles
with momenta between 20 and 55~GeV/\sol\ as protons is less than 3\%. 
In the \ks \ sample the \lam\ and \alam\ contaminations were removed \cite{zav03}. 
In addition, only p\ks\ pairs from events with an identified 
primary vertex were accepted.  The invariant mass spectrum of the p\ks\ pairs  is shown in 
Figure \ref{d2}a) for the p+C data.
The spectrum exhibits a smooth shape without any narrow structure. The solid line represents the
background determined from event mixing. In the background-subtracted spectrum (Figure \ref{d2}b),
the data points scatter around zero showing again no evidence for a narrow structure in the mass
region between 1450 and 1700 MeV/\solq . From Monte-Carlo simulations, the experimental mass resolution 
for a \thp\ is found to be $\sigma \sim 3.8$~MeV/\solq .
The estimated sensitivity 
{\it Br}$\times$d$\sigma$/dx$_F$ for a narrow signal is about 5~$\mu$b~/~nucleon assuming the
cross section to scale with A$^{0.7}$. 
Also other search strategies did not yield a significant narrow structure around 1540~MeV/\solq\ 
including 
i) a cut in the track multiplicity of the event (Figure \ref{d2}c), 
ii) the request for the presence of a tagging particle like \lam , $\Sigma$ or \km\ in the event, or 
iii) both conditions (Figure \ref{d2}d), and
iv) the relaxation of the proton momentum cut implied by the RICH PID.
With the same proton PID used for the p\ks\ pairs, the combination of protons with \km\ 
particles yields a clean signal of the \lam(1520)~$\rightarrow$~p\km\ decay (Figure \ref{d1}e) 
proving  the PID performance. 
In fact, \thp~$\rightarrow$~p\ks\ and \lam(1520)~$\rightarrow$~p\km\ decays
exhibit a very similar geometrical acceptance and, assuming a branching ratio of 0.25 of the \thp\ into
the p\ks\ channel, a preliminary  value for the upper limit of the relative particle yield 
\thp/\lam(1520) at mid-rapidity is 0.02 at 95\% CL. This value is significantly lower 
than the statistical hadronization model prediction of 0.57 for p-p collisions 
at $\sqrt{s} = 17$~GeV \cite{bec03}.

\section{Search for \ximm~$\rightarrow$~\xim~+~\pim\  decays}
\label{sec:searchxi}

\begin{figure}[h]
\centering
\includegraphics[width=14cm]{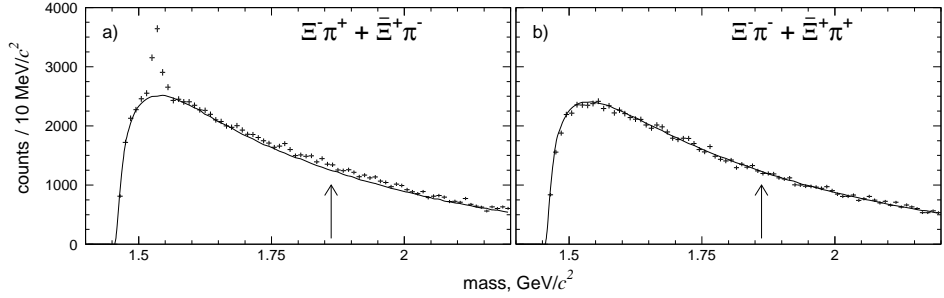}
\caption{\label{d3} 
The $\Xi\pi$ invariant mass distributions obtained with all targets C, Ti, and W in indicated
    decay channels. Continuous lines show the background from event mixing. Arrows mark the
    mass of 1862~MeV/\solq .
}
\end{figure}

Like in the NA49 study \cite{na49}, resonances were searched for in both the doubly-charged 
\xim\pim +\,c.c.  and in the neutral  \xim\pip +\,c.c. channels. 
$\Xi$ candidates with a mass of $\pm 10$~MeV/\solq\ of the table mass were accepted. 
Figure \ref{d3} shows the 
corresponding invariant mass spectra and the backgrounds determined by event mixing.
In the neutral decay channels, (Figure \ref{d3}a), the \xisz\ resonance shows up  with a prominent 
signal, and there is a possible weak evidence for known higher \xis\ resonances. In the doubly-charged
channels (Figure \ref{d3}b), the background follows very well the data, and, in particular, there is 
no evidence for a narrow signal at around 1862~MeV/\solq\  at a sensitivity of 
{\it Br}$\times$d$\sigma$/dx$_F|_{x_F\sim 0}  < 25~\mu$b~/~nucleon. 
The preliminary values for the two upper
limits of the relative production yields Br$\times$\ximm(1862)/\xisz\ and Br$\times$\xipp(1862)/\axisz\ at
mid-rapidity are 0.077 and 0.058 at 95\% CL, respectively.

\section{Conclusions}
\label{sec:concl}

The present study exploited a high statistics minimum bias sample obtained from central proton 
nucleus collisions at $\sqrt{s} = 41.6$~GeV and the high momentum resolution of the HERA-B
spectrometer for a sensitive search of narrow pentaquark signals.   
The performance of the analysis technique has been verified by the reconstruction of clear 
\lam(1520)~$\rightarrow$~p\km\ and  \xisz $\rightarrow$~\xim\pip\ signals. 
No evidence is found, however, for narrow signals from the \thp\ around 1540~MeV/\solq\ 
or the \ximm\ candidate at about 1862~MeV/\solq\ in the p\ks\ and \xim\pim channels, respectively. 
The sensitivity to identify a \thp\ signal in the p\ks\ channel is estimated to be better than 
5~$\mu$b~/~nucleon. 
More systematic studies and the evaluation of cross section limits are in progress. 
The relative particle yield \thp /\lam(1520) of less than 0.02 (95\% CL) at mid-rapidity is not 
compatible with the predictions of the statistical hadronization model \cite{bec03}.
If existent, pentaquarks seem to exhibit also exotic production mechanisms.


\section*{References}

\begin{thebibliography}{99}


\bibitem{leps} Nakano T \etal (LEPS collaboration) 2003 \PRL {\bf 91} 012002 
\nonum         Stepanyan S \etal (CLAS collaboration) 2003 \prep\ hep-ex/0307018
\nonum         Bart J \etal (SAPHIR collaboration) 2003 \PL {\bf 572} 127
\nonum         Barmin V V \etal   (DIANA collaboration)  2003 Yad. Fiz. {\bf 66} 1763 
\nonum         Asratyan A E \etal (ITEP collaboration)   2003 \prep\ hep-ex/0309042
\nonum         Kubarovsky V \etal (CLAS collaboration)  2003 \prep\  hep-ex/033046
\nonum         Airapetian A \etal (HERMES collab.) 2003 \prep\ hep-ex/0312044
\bibitem{cah03} Cahn R N and Trilling G H 2003 \prep\ hep-ph/0311245 and references therein
\bibitem{dia97} Diakonov D, Petrov V and Polyakov M V 1997 \ZP A {\bf 359}  305
\bibitem{jaf03} Jaffe R and Wilczek F 2003 \prep\ hep-ph/0307341 and proceedings of this conference
\bibitem{na49} Alt C \etal (NA49 collaboration)   2003 \prep\ hep-ex/0310014
\bibitem{ran03} Randrup J 2003 \PR C {\bf 68} 031903 
\bibitem{let03} Letessier J \etal 2003   \prep\ hep-ph/0310188
\bibitem{bec03} Becattini \etal 2003     \prep\ hep-ph/0310049
\bibitem{abt03}  Abt I \etal (HERA-B collaboration) 2003 \EJP C {\bf 29} 181-190 and references therein  
\bibitem{zav03}  Zavertyaev M 2003 \prep\ hep-ph/0311250

\end{thebibliography}
\end{document}